\documentclass[aps, preprint, 12pt]{article}
\usepackage{times}
\usepackage{amssymb}
\usepackage{amsmath}
\usepackage{array}
\usepackage{float}
\usepackage[margin=2.5cm]{geometry}
\usepackage{fancyhdr}
\usepackage{cite}
\usepackage[format=plain,=small,labelfont=bf,labelsep=period]{caption}
\usepackage{authblk}

\usepackage{color}
\usepackage{pdfcolmk} 
\usepackage{setspace}
\usepackage{soul}
\usepackage[hidelinks]{hyperref}

\usepackage{pdfpages}    

\ifx\pdfoutput\undefined
\usepackage{graphicx}
\else

\usepackage{pdflscape} 

\usepackage[version=3]{mhchem} 

\newcolumntype{x}[1]{>{\raggedright\hspace{0pt}}p{#1}}

\graphicspath{{.},{./f/}}

\usepackage{threeparttable}

\title{Revolutionizing Alloy Microstructure Segmentation through SAM and Domain Knowledge without Extra Training}

\author{Xudong Ma$^{\text{1}}$, Yuqi Zhang$^{\text{1}}$, Chenchong Wang$^{\text{1},~\text{*}}$, Wei Xu$^{\text{1}}$}
\affil{$^\text{1}$ State Key Laboratory of Rolling and Automation, Northeastern University, Shenyang, Liaoning 110819, China \\$\text{*}$ Corresponding authors: Email: wangchenchong@ral.neu.edu.cn, Tel: +024-83680246\\}

\begin{document}
\date{}
\maketitle


\begin{abstract}
\noindent
Fundamental models, trained on large-scale datasets and adapted to new data using innovative learning methods, have revolutionized various fields. In materials science, microstructure image segmentation plays a pivotal role in understanding alloy properties. However, conventional supervised modelling algorithms often necessitate extensive annotations and intricate optimization procedures. The segmentation anything model (SAM) introduces a fresh perspective. By combining SAM with domain knowledge, we propose a novel generalized algorithm for alloy image segmentation. This algorithm can process batches of images across diverse alloy systems without requiring training or annotations. Furthermore, it achieves segmentation accuracy comparable to that of supervised models and robustly handles complex phase distributions in various alloy images, regardless of data volume. \\
\\
{\singlespace \footnotesize \noindent \textbf{Keywords:} Alloy material, Fundamental model, Domain knowledge, Segmentation, Post-processing}
\end{abstract}





\newpage

\section{Introduction}
Pre-trained large language models on web-scale datasets represent a burgeoning field, featuring innovative learning methods that adapt to new data across diverse environments with minimal additional training (i.e., few-shot~\cite{RN359} or zero-shot learning~\cite{RN357}). In natural language processing~\cite{RN382, RN385}, these foundational models have become the predominant force, gradually supplanting traditional task-specific methods and catalysing paradigm shifts in other domains like computer vision~\cite{RN368,RN369,RN370}.

Recently, the segment anything model (SAM)~\cite{RN371} has garnered considerable attention as an innovative foundational model for image segmentation. This model has been trained on a dataset comprising over 1.1 billion segmentation masks. It employs an interactive learning method and demonstrates robust zero-shot segmentation performance across various natural image datasets.


The core paradigm of alloy design is to understand and utilize composition/processing-structure-property relationships. 
Microscopic image segmentation is often the first and critical step in quantifying the structure of alloys~\cite{RN345}. 
Task-specific supervised models have shown substantial results~\cite{RN346, RN347, RN349, RN350, RN353, RN354}, but their generalization capabilities are poor. SAM offers a potential 
solution, yet its effectiveness across various microstructural morphologies and task requirements lacks comprehensive evaluation. 
Furthermore, the highly subjective nature of SAM’s interactive 
process impedes rapid and uniform processing of batch data.

In this study, we develop a novel alloy image segmentation technique grounded in SAM and domain knowledge. This technique, which requires no manual annotation or model debugging, can be applied across various alloy systems. The process involves obtaining initial segmentation results from SAM and subsequently post-processing them using domain knowledge. This approach achieves segmentation accuracies comparable to those of supervised learning algorithms, while significantly minimizes the need for manual intervention, offering a fresh approach to quantifying the microstructure of alloy materials.

\section{Datesets and methods}

\subsection{Generalized segmentation method for alloy images}
\begin{figure}[h]
\centering
\includegraphics[width=6.5in]{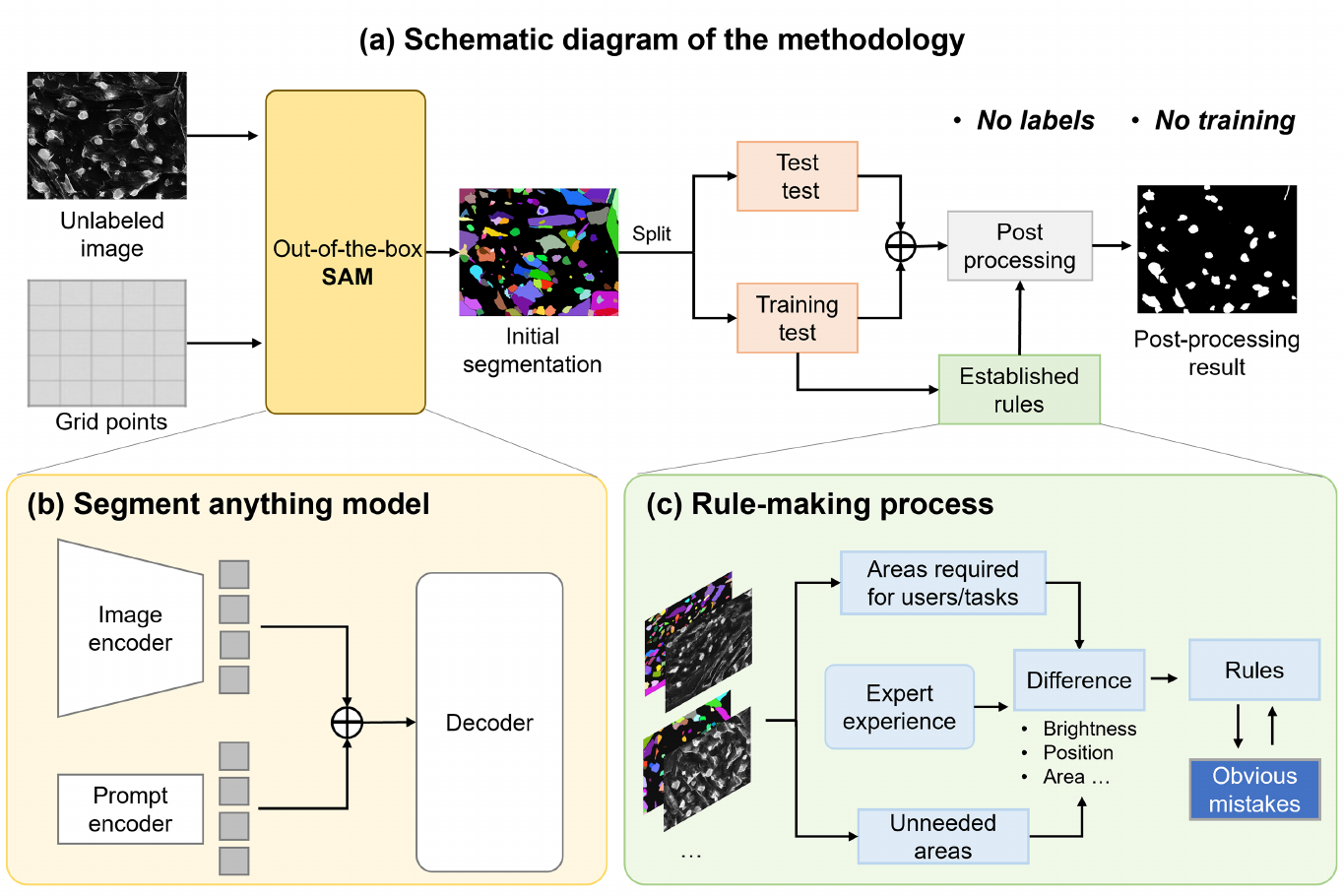}
\caption{Overview of the methodology. (a) Schematic diagram of the methodology. (b) Structure of the segment anything model (SAM). (c) Rule-making process. }
\label{fig:method}
\end{figure}

Fig.~\ref{fig:method} depicts a fresh semantic segmentation method. This algorithm comprises an initial SAM-based segmentation phase and a subsequent post-processing phase driven by expert experience. Specifically, SAM incorporates three primary components, as illustrated in Fig.~\ref{fig:method}(b): an image encoder, a prompt encoder, and a mask decoder. The image encoder employs a visual transformer~\cite{RN374} as its architecture and is pre-trained using a mask autoencoder~\cite{RN372} strategy. The prompt encoder transforms various prompts into vector representations, while the mask decoder converts all embeddings into masks. To avoid the impact on accuracy from the cueing method, SAM’s ‘automatic mask generation’ function is utilized (use grid points as prompts), with SAM operating under its default architecture and parameters. The initial SAM-based segmentation results encompass both required and superfluous regions, necessitating rule-based post-processing.

As illustrated in Fig.~\ref{fig:method}(c), in accordance with task requirements, we view the identified regions within the training set images and select simple metrics, such as brightness. Utilizing these metrics, we formulate a set of rules to be implemented on the training set images and rectify these rules for instances of error. Ultimately, these rules are applied across the entire dataset.

\subsection{Overview of the datasets}
We assessed the efficacy of the current method in alloy segmentation tasks utilizing four classical datasets, each representing different microstructure morphologies. The first dataset, sourced from ultra-high carbon steel (UHCS)~\cite{RN354}, comprises 17 low-magnification images. These images involve a binary classification of spheroidite particles and matrix, serving to validate the segmentation capability of the current method for images featuring small and densely packed particles. The second dataset, referred to as the carbide dataset and derived from the authors’ previous work~\cite{RN353}, consists of 26 low-magnification images of hot-formed steels. These images are labelled with carbides and matrix, and this dataset is difficult to segment due to the presence of numerous bright boundary interferences. The third dataset pertains to nickel-based superalloys (hereafter referred to as Super)~\cite{RN345} and involves a three-classification task encompassing the matrix phase, secondary precipitates, and tertiary precipitates. The fourth dataset represents a binary classification task for environmental barrier coatings (EBC)~\cite{RN345}, covering both oxide layers and matrix. Detailed information on the datasets is provided in the supplementary materials, as shown in Fig.~\ref{fig: data images} and Table~\ref{data table}.

\section{Results and discussion}
\subsection{Initial segmentation results}

\begin{figure}[h]
\centering
\includegraphics[width=6.5in]{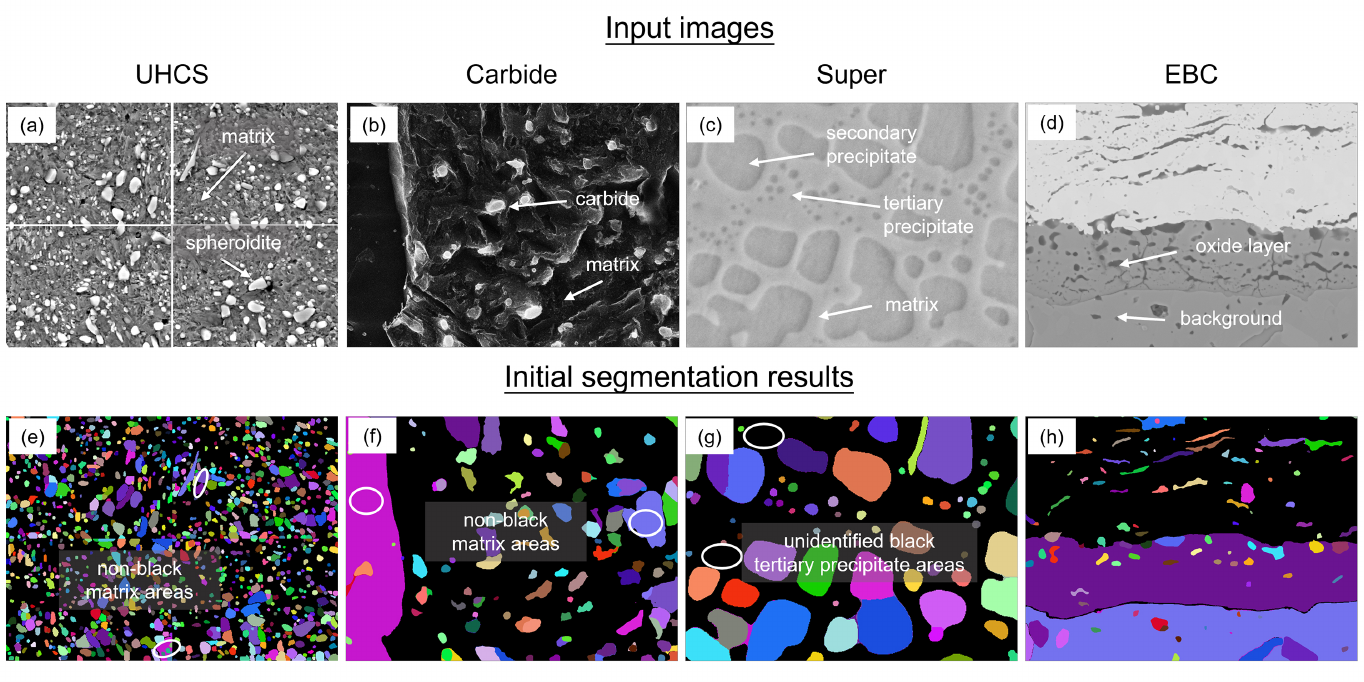}
\caption{Initial segmentation results. Representative images of test sets from (a) ultra-high carbon steels (UHCS), (b) carbide dataset, (c) nickel-based superalloys (subsequently referred to as Super), and (d) environmental barrier coating (EBC). (e)-(h) Corresponding initial segmentation results from various datasets.}
\label{fig:initial results}
\end{figure}

Fig.~\ref{fig:initial results} presents the initial segmentation results for the test set across various datasets. For the UHCS image, we divided the original image into four equal segments, input these into SAM, and subsequently combined the results to obtain a complete segmentation map. As depicted in Fig.~\ref{fig:initial results}(e), the matrix is represented in black, with most particles accurately identified and assigned distinct colours. However, certain elevated or depressed matrices were also represented as non-black due to varying degrees of corrosion, complicating the distinction between matrix and spheroidite particles based solely on colour. The carbide dataset encounters a similar issue, as shown in Fig.~\ref{fig:initial results}(f). Fig.~\ref{fig:initial results}(g) and (h) display the identification results for the Super and EBC datasets, which is overall accurate, but there is also no discernible pattern between the colours of the matrix and oxide layer in the EBC. Consequently, the images cannot be classified merely by colour differences, necessitating the formulation of specific rules to post-process the initial SAM-based segmentation results.

\subsection{Domain knowledge-driven post-processing}

\begin{figure}[h]
\centering
\includegraphics[width=6.5in]{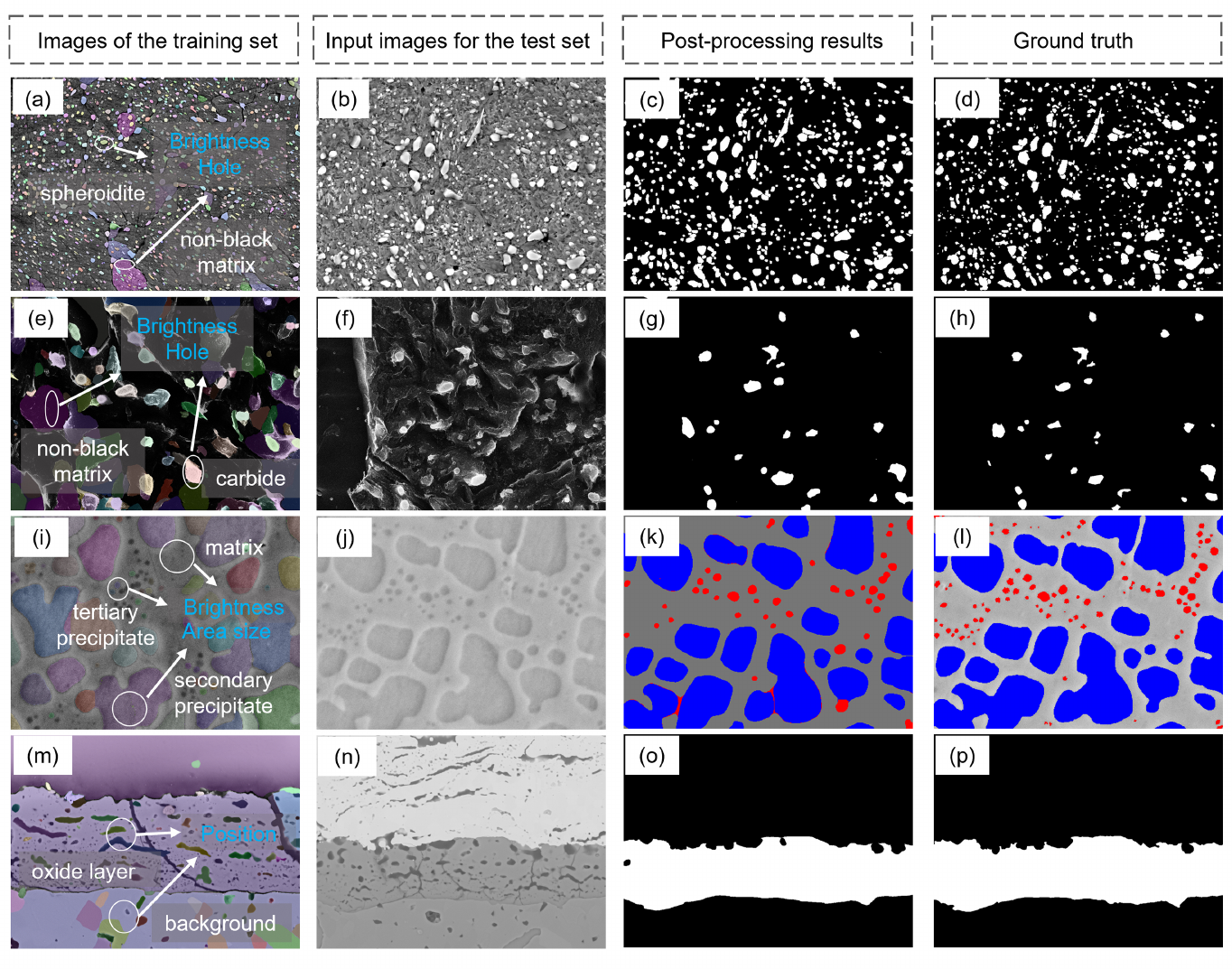}
\caption{Formulation of post-processing rules and their results. Overlay results of training set images and initial segmentation maps for (a) UHCS, (e) carbide, (i) Super, and (m) EBC. (b), (f), (j), and (n) represent sample images from the test set. (c), (g), (k) and (o) describe the corresponding post-processing results of the sample images of the test set after the first amendment to the rules. (d), (h), (l), and (p) are the corresponding labelled images.}
\label{fig:post-processing results}
\end{figure}

In both carbide and UHCS, we interpret the black areas of the initial segmentation results as matrix, while the non-black areas may represent either matrix or precipitates. As illustrated in Fig.~\ref{fig:post-processing results}(a) and (e), the distinction between the non-black matrix and non-black precipitates primarily lies in the presence or absence of holes and the average brightness. Consequently, the presence of holes is evaluated first, followed by the application of a brightness threshold to differentiate the two, with areas exhibiting holes or brightness below the threshold classified as the matrix. The process of setting the threshold is as follows: initially, the average brightness of images is standardized to a fixed value to eliminate the brightness difference. The fixed value is determined as the minimum of the average brightness of the training set to prevent brightness overflow (the brightness range is between 0 and 255). Subsequently, three training set samples are selected to generate their reference images based on their initial results by eliminating unwanted colours. The brightness range is then defined based on the non-black regions of the training set, and the spacing is calculated using Sturges’ formula~\cite{RN379}. Post-processing is conducted on each sample from minimum to maximum brightness with fixed spacing, semantic segmentation metrics are calculated based on the reference images (indicators are consistent with those of the paper from which the data were obtained) and monitored for changes, with the optimal brightness value subsequently determined. As shown in Fig.~\ref{fig: trends} in the supplementary materials, the metrics generally increased and then decreased; if there was a monotonic change, the spacing is adjusted to half, a third quarter of the original, and so forth, until the trend was correct. Ultimately, the average of the three tests is adopted as the threshold value. The thresholds for the UHCS and carbide datasets were 108 and 61, respectively. As depicted in Fig.~\ref{fig:post-processing results}(c) and (g), the particle morphologies of the test set are largely consistent with the labelled images.

In Super alloys, the matrices typically correspond to the black areas in the initial segmentation results, although a few precipitated phases may also be identified as black. As illustrated in Fig.~\ref{fig:post-processing results}(i), the secondary and tertiary precipitates are generally less bright than the matrix, and both differ in area. Consequently, we utilize the average brightness of the part of the original image corresponding to the black region with the largest area in the initial segmentation result as a standard, with brightness levels below this standard recognized as precipitates and those above it as matrix. The secondary and tertiary precipitates are differentiated by the size of the area and the threshold is determined by a process akin to that for the carbide dataset, except that the brightness is substituted for the area and the area doesn’t require standardization. Because of insufficient scale information, we used pixels as a metric and set the area threshold to 585 pixels. The results, depicted in Fig.~\ref{fig:post-processing results}(k), generally exhibit accurate phase division, although there is some under-segmentation in the tertiary precipitations. For EBC alloy, as shown in Fig.~\ref{fig:post-processing results}(m), the primary distinction between the oxide layer and the background area is the location. If we set the image height as h, establish the lower left corner of the image as the origin, and establish the coordinate axes, the y-values of all the points of the oxide layer should be greater than $\alpha$ pixels and less than h-$\alpha$ pixels, and $\alpha$ is initially set to 0. Since the area of the oxidized layer is relatively large, we arrange the areas in the initial segmentation results by area from largest to smallest, and look for the first area that meets the above conditions, which is recognized as the oxide layer, and conduct the filling operation to reduce the influence of impurities on the recognition accuracy. The results of the training set are depicted in Fig.~\ref{fig: rule correction} of the supplementary materials, where some recognition results appear to be unreasonable. Specifically, as illustrated in Fig.~\ref{fig: rule correction}(g), the region near the image’s edge is erroneously identified as an oxide layer. Furthermore, as demonstrated in Fig.~\ref{fig: rule correction}(c) and (k), the recognized area of the oxidized layer is too short. To address this issue, additional rules were added as follows, called the first amendment to the rules: 1. $\alpha$ was reset to 10; 2. if the width of the region of the recognized oxide layer is less than half of the image, or if there is no region that satisfies all of the above conditions, the remainder of the region with the largest and the second largest area is considered as the oxide layer. The results are depicted in Fig.~\ref{fig: rule correction}(d), (h) and (l) of the supplementary materials and Fig.~\ref{fig:post-processing results}(o), and the outline of the oxide layer in the test set is clearly visible.

\subsection{Comparison of the current methodology with the supervisory modeling approach}

\begin{figure}[htbp]
\centering
\includegraphics[width=6.5in]{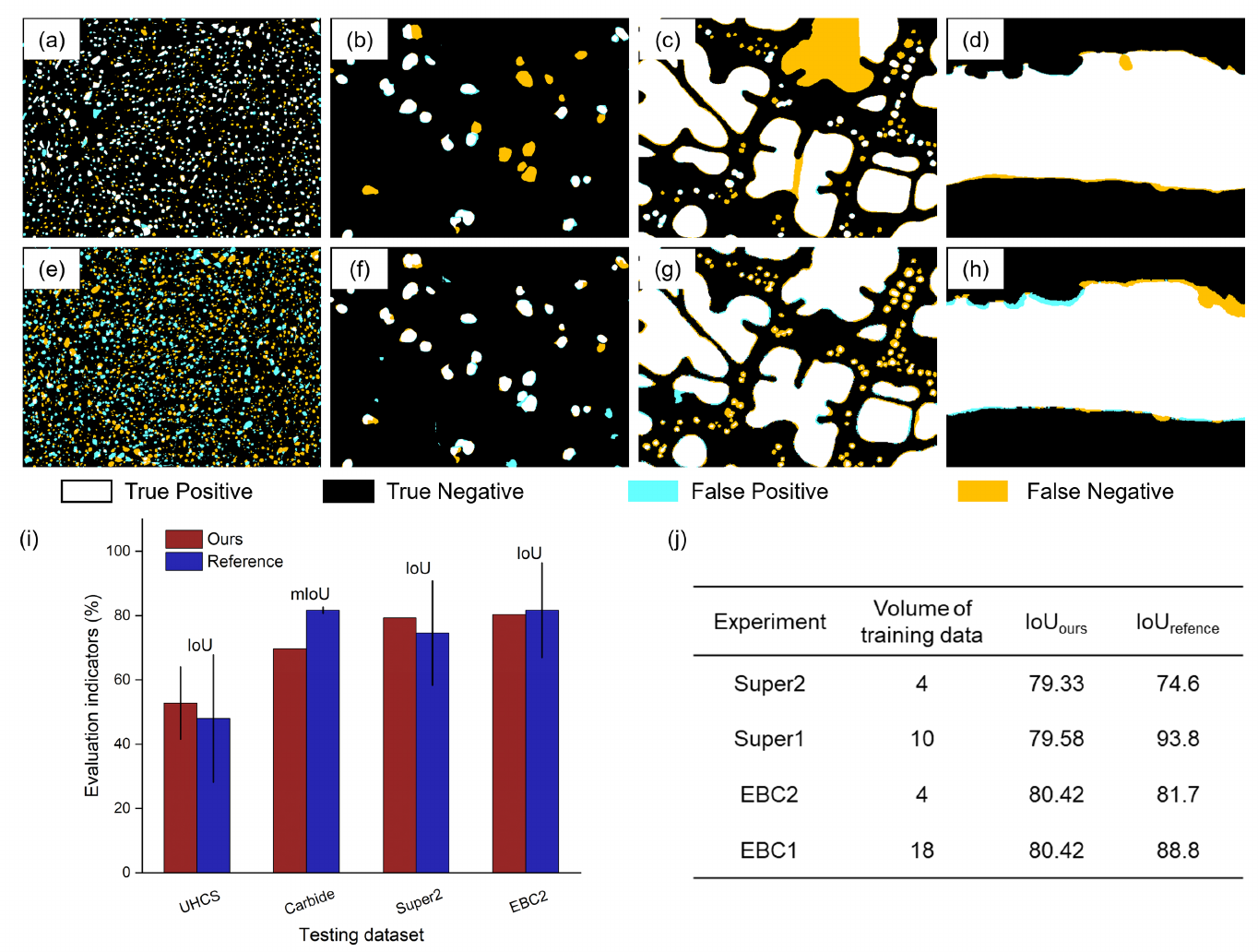}
\caption{Comparison of the results of the current method with the supervised model on the testing set. Super2 and EBC2 are the datasets used above, while Super1 and EBC1 are datasets with larger training data volumes, including Super2 and EBC2, respectively. Detailed information on the datasets is provided in the supplementary materials, as shown in Fig.~\ref{fig: data images} and Table~\ref{data table}. Segmentation results of the current method for representative images in the (a) UHCS, (b) Carbide, (c) Super2, and (d) EBC2. (e)-(h) Corresponding segmentation results from the supervised model. (i) Comparison of evaluation metrics between the two methods. For UHCS using cross-validation, we average prediction results of all images to represent the final test set results. The evaluation metrics employ intersection over union (IoU) or mean intersection over union (mIoU), and the selection rules are based on the literature on supervised models. White denotes true positives, black indicates true negatives, cyan signifies false positives, and orange symbolizes false negatives. (j)  The variation of evaluation metrics of Super and EBC testing sets with varying amounts of training data.  The rule or threshold transformations induced by the amount of training data are shown in Note 1 of the supplementary materials.}
\label{fig: comparison}
\end{figure}

Fig.~\ref{fig: comparison} compares the segmentation results of the current method with those of the supervised model across various datasets. The intersection over union (IoU) performance of the current method is comparable to that of the supervised model in the Super, EBC, and UHCS datasets. Notably, for UHCS, where cross-validation is employed by the supervised model, we present an average of prediction results of all images as representative of the performance of test set. 

As depicted in Fig.~\ref{fig: comparison}(a), (e), and (i), our method significantly reduces both false-positive and false-negative regions within UHCS, significantly surpassing the supervised model’s performance.This improvement stems from the robust prediction capabilities of SAM and the application of simple techniques such as segmentation. For the carbide test set, as shown in Fig.~\ref{fig: comparison}(b), (f), and (i), despite the differences in the mean intersection over union (mIoU) between the current method and the supervised model, there is minimal difference in the size of the false-positive and false-negative regions between the two. The slightly higher number of false-negative regions in the current method may be associated with the brightness threshold being set too high. Accordingly, lowering the threshold appropriately may serve as an effective strategy to enhance recognition accuracy.

To highlight the benefits of current approach, we examined the influence of the volume of training data on the accuracy of segmentation. The quantity of training samples was augmented from 4 to 10 and 18 for the Super and EBC datasets, respectively. Fig.~\ref{fig: different data volumes} and Fig.~\ref{fig: comparison}(j) reveal that the quantity of training data profoundly impacts the supervised model, and the accuracy of the supervised model on the test set escalates significantly with the increase in the quantity of training data. In contrast, the accuracy of current approach remains largely stable, which demonstrates the robustness of our approach, and also indicates that current approach can achieve optimal results even with a minimal volume of data.

\section{Conclusion}
In this study, we have successfully developed a generalised automatic segmentation algorithm that combines a large language model with domain knowledge. This represents the first unified microstructural segmentation endeavour to achieve zero training for alloy materials. Specifically, the research offers three main contributions. First, the traditional data-driven training paradigm for microstructure segmentation is broken. Second, we demonstrate that the formulation of uniform rules through expert experience can be effectively adapted to meet different segmentation task requirements. Finally, the results demonstrate that the current method is comparable to supervised models in terms of segmentation accuracy. Nevertheless, this study still faces some challenges in practical applications, such as the need for experts to customise the rules for different datasets. We hope that this study can provide some new insights for quantitative analysis of alloy materials.

\subsection*{Acknowledgments}

The research was supported by the National Key Research and Development Program of China (No. 2022YFB3707501), the National Natural Science Foundation of China (No. U22A20106 and No. 52304392). 



\subsection*{Author contributions}
X.M., Y.Z., and C.W. developed the idea and conducted the research. X.M. and Y.Z. were responsible for data collection, data cleaning, and model conceptualization (including design architecture, code implementation, training, and evaluation). X.M. and C.W. performed the validation. X.M., and Y.Z. wrote the manuscript. W.X., C.W., and Y.Z. provided guidance for the study. W.X. provided financial support. All authors reviewed the final manuscript.

\subsection*{Data availability}
The data that support the findings of this study are available from the corresponding author upon reasonable request.

\subsection*{Code availability}
The codes are available from the corresponding author upon reasonable request.


\subsection*{Declaration of competing interest}

The authors declare that they have no known competing financial interests or personal relationships that could have appeared to influence the work reported in this paper.


\newpage

\begin{center}
	Supplemental Materials for\\[4mm]
	{\Large\textbf{Revolutionizing Alloy Microstructure Segmentation through SAM and Domain Knowledge without Extra Training}}\\[4mm]
Xudong Ma$ ^1 $, Yuqi Zhang$ ^1 $, Chenchong Wang$ ^{1,~*} $, Wei Xu$ ^1 $\\[2mm]
\textit{$ ^{\textit{1}} $ State Key Laboratory of Rolling and Automation, Northeastern University, Shenyang, Liaoning 110819, China}\\[2mm]
* Corresponding Author E-mail address: wangchenchong@ral.neu.edu.cn
\end{center}

\vspace*{1cm}
{\large \textbf{This supplementary material includes:}}
\begin{table}[htpb]
\centering
\large 
\begin{tabular}{p{14cm}}
\textbf{Fig. \ref{fig: data images}.} Representative images from the dataset.\\
\\
\textbf{Fig. \ref{fig: trends}.} Segmentation evaluation metric trends for UHCS, Carbide, and Super2 sample images as brightness or area varies from maximum to minimum with fixed spacing.\\
\\
\textbf{Fig. \ref{fig: rule correction}.} Failure cases and amended results for the EBC1 and EBC2 training sets.\\
\\
\textbf{Fig. \ref{fig: different data volumes}.} Variation of segmentation accuracy of current method with different amount of training data.\\
\\
\textbf{Table \ref{data table}.} Number of images in the train and test splits of each experimental dataset.\\
\\
\textbf{Supplementary Note 1}\\
\end{tabular}
\end{table}

\newpage

\renewcommand\thefigure{S\arabic{figure}}
\setcounter{figure}{0}

\begin{figure}[H]
\centering
\includegraphics[width=6.5in]{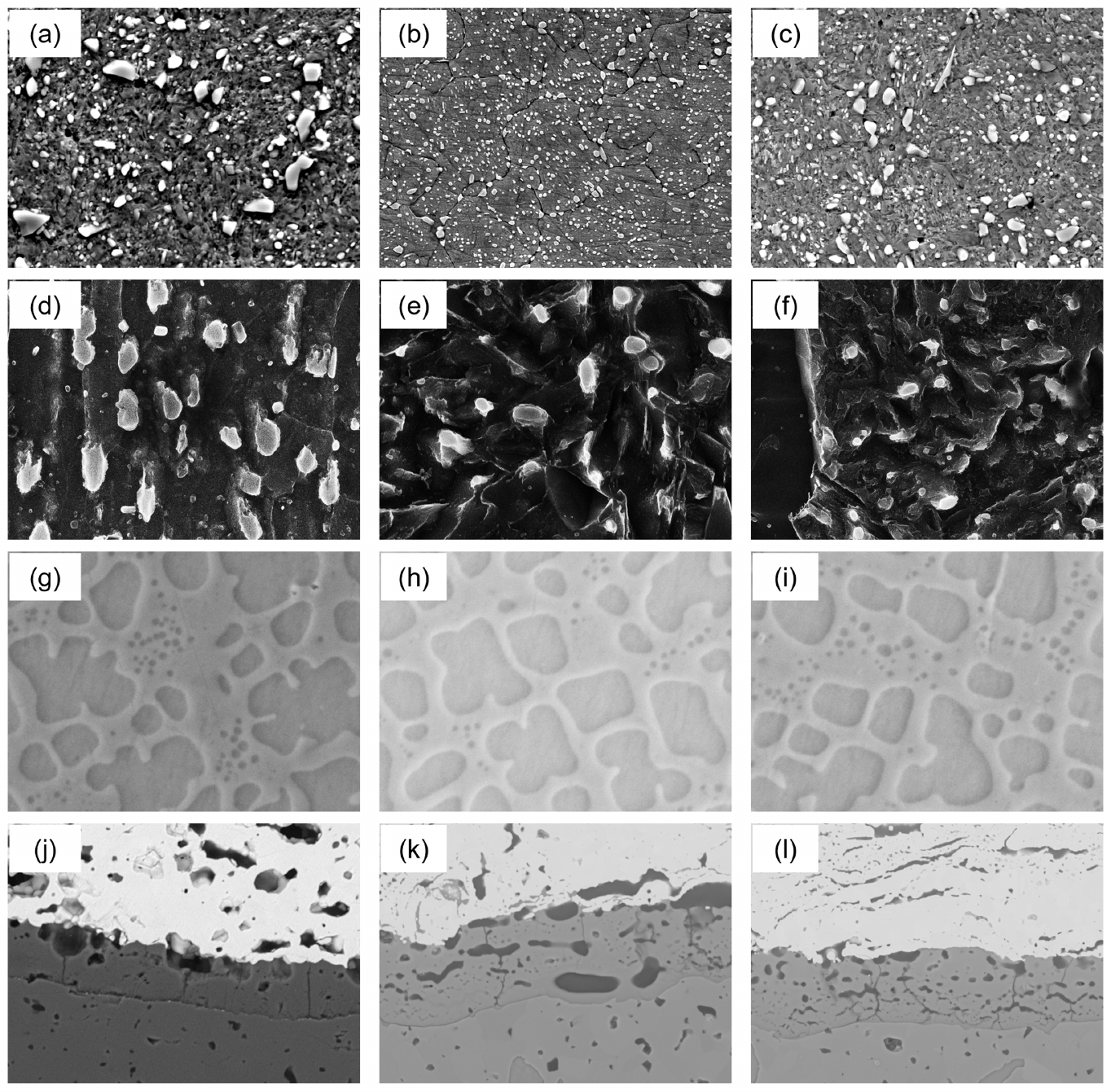}
\caption{Representative images from the dataset. (a)-(c) Ultra-high carbon steel (UHCS) dataset. (d)-(f) carbide dataset. (g)-(i) Super dataset. (j)-(l) EBC dataset.}
\label{fig: data images}
\end{figure}

\begin{figure}[H]
\centering
\includegraphics[width=6.5in]{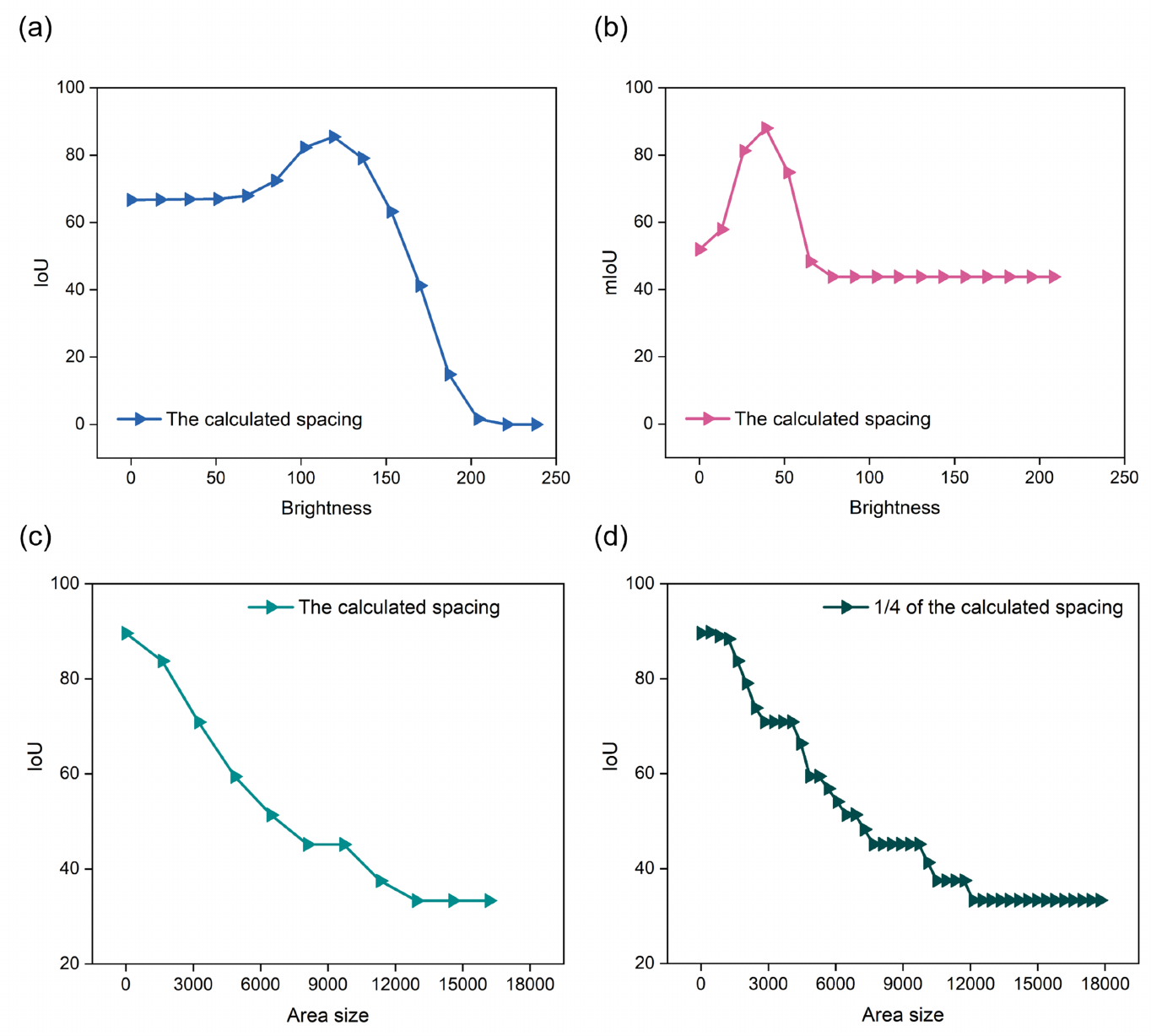}
\caption{Segmentation evaluation metric trends for UHCS, Carbide, and Super2 sample images as brightness or area varies from maximum to minimum with fixed spacing. (a)-(b) Results of UHCS and carbide dataset with spacing determined by Sturgis’ rule~\cite{RN379}. (c)-(d) Results of Super2 dataset at full and quarter spacing, as determined by Sturgis’ rule~\cite{RN379}.}
\label{fig: trends}
\end{figure}

\begin{figure}[H]
\centering
\includegraphics[width=6.5in]{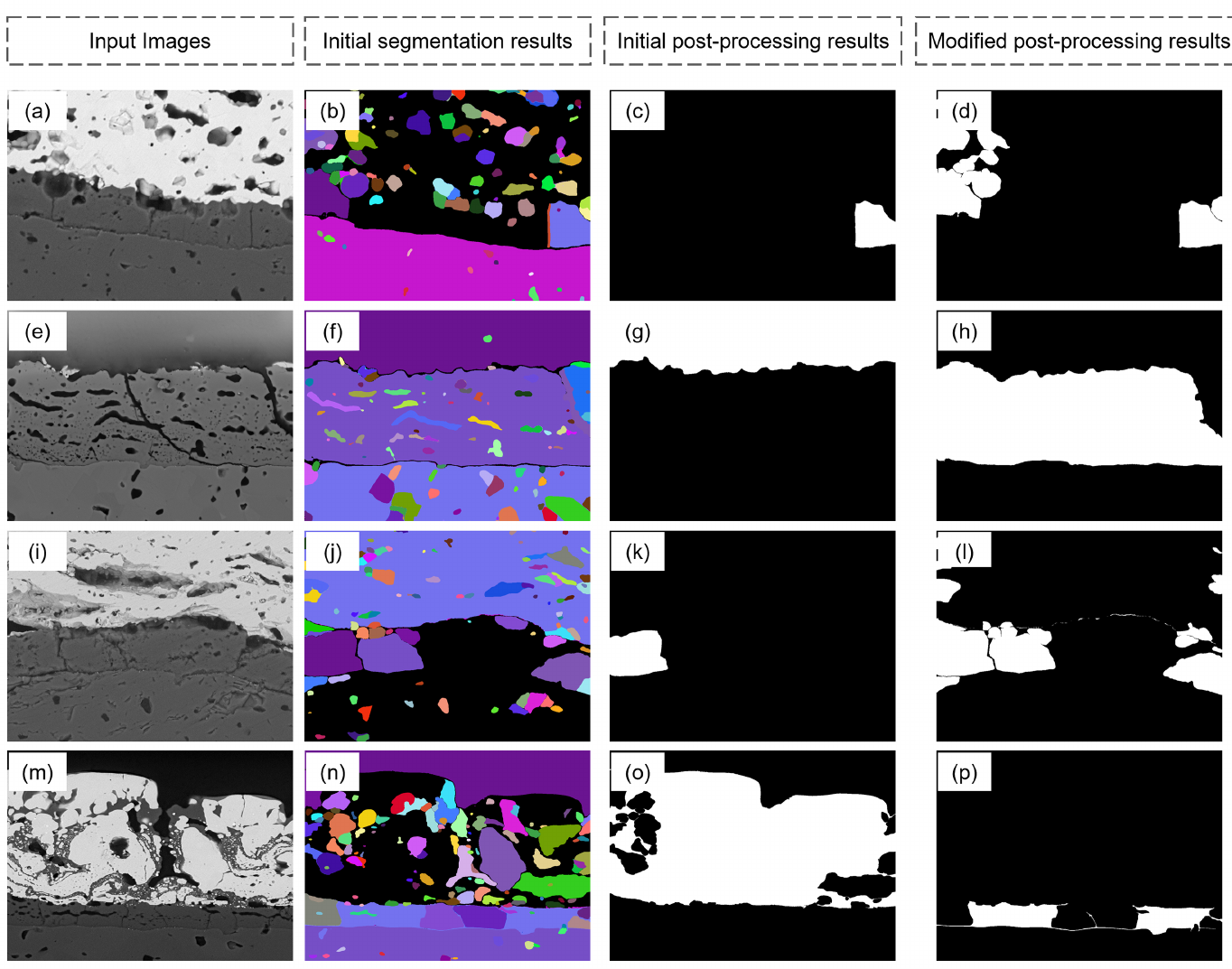}
\caption{Failure cases and amended results for the EBC1 and EBC2 training sets. (a), (e), (i), and (m) display representative images from the training set. (b), (f), (j), and (n) present the corresponding initial segmentation results. (c), (g), and (k) show the post-processing results of the EBC2 dataset based on the initial rules. (d), (h), and (l) exhibit the results of the EBC2 dataset after the first amendment of the rules. Given that the EBC1 dataset encompasses the EBC2 dataset, the first amended rules are directly applied to the EBC1 dataset. (o) displays the new failure cases, so we introduce an additional new rule stating that the height of the oxidized layer exceeds 50 pixels but does not surpass half of the image height, which is called the second amendment of the rules. The final result is shown in (p) and the recognized oxide layer is more reasonable.}
\label{fig: rule correction}
\end{figure}

\begin{figure}[H]
\centering
\includegraphics[width=6.5in]{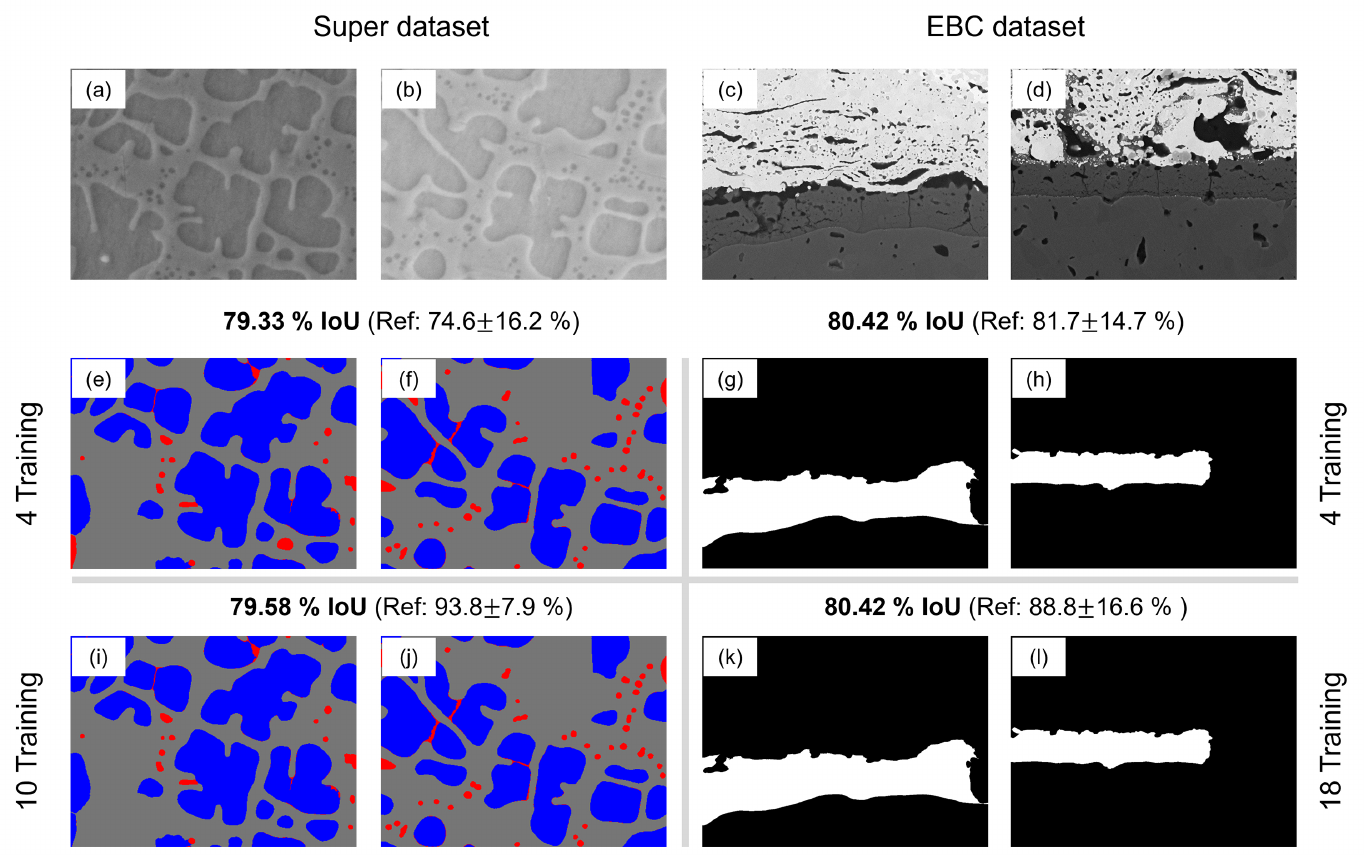}
\caption{Variation of segmentation accuracy of current method with different amount of training data. (a)-(d) Representative samples from the test sets. (e)-(h) Segmentation outcomes and corresponding accuracy for the EBC2 and Super2 datasets. (i)-(l) Segmentation outcomes and corresponding accuracy for the EBC1 and Super1 datasets.}
\label{fig: different data volumes}
\end{figure}

\renewcommand\thetable{S\arabic{table}}
\setcounter{table}{0}

\begin{table}[h]
    \setlength{\tabcolsep}{60pt}
    \centering
    \caption{Number of images in the train and test splits of each experimental dataset.}
    \begin{threeparttable}
    \begin{tabular}{ccc}
    \hline
        Experiment & \# Train & \# Test \\ \hline
        UHCS & 14 & 3 \\ 
        Carbide & 59 & 5 \\ 
        Super2 & 4 & 4 \\ 
        Super1 & 10 & 4 \\ 
        EBC2 & 4 & 3 \\ 
        EBC1 & 18 & 3 \\ \hline
    \end{tabular}

    \begin{tablenotes}
    \footnotesize
     \item[*] The Super2 dataset from nickel-based superalloys (hereinafter referred to as "Super") and the EBC2 dataset from environmental barrier coating (EBC) alloys are used in Figs.~\ref{fig:initial results}-\ref{fig: comparison} of the main text. Super1 and EBC1 are datasets with larger training data volumes, including Super2 and EBC2, respectively.  
    \end{tablenotes}
    \end{threeparttable}
    \label{data table}
    
\end{table}

\noindent
{\large\textbf{Supplementary Note 1}}

\noindent
The Super2 dataset used in Figs.~\ref{fig:initial results}-\ref{fig: comparison} of the main text employs an area threshold of 585 pixels, while the Super1 dataset uses a threshold of 495 pixels. Contrasting with the EBC2 dataset used in Figs.~\ref{fig:initial results}-\ref{fig: comparison} of the main text, the EBC1 dataset introduces an additional criterion: the recognized oxide layer’s height must be greater than 50 pixels but less than half the image’s total height.

\end{document}